\documentclass[superscriptaddress,showpacs,twocolumn]{revtex4}
\usepackage{amssymb}
\usepackage{amsmath}
\usepackage{graphicx}
\usepackage{epsfig}

\begin{document}

\title{Quantum transducers: Integrating Transmission Lines and Nanomechanical Resonators via Charge
Qubits}
\author{C. P. Sun}
\affiliation{Frontier Research System, The Institute of Physical and Chemical Research
(RIKEN), Wako-shi, Saitama, 351-0198, Japan}
\affiliation{Institute of Theoretical Physics, The Chinese Academy of Sciences, Beijing,
100080, China}
\author{L. F. Wei}
\affiliation{Frontier Research System, The Institute of Physical and Chemical Research
(RIKEN), Wako-shi, Saitama, 351-0198, Japan}
\affiliation{Institute of Quantum Optics and Quantum Information, Department of Physics,
Shanghai Jiaotong University, Shanghai 200030, China}
\author{Yu-xi Liu}
\affiliation{Frontier Research System, The Institute of Physical and Chemical Research
(RIKEN), Wako-shi, Saitama, 351-0198, Japan}
\author{Franco Nori}
\affiliation{Frontier Research System, The Institute of Physical
and Chemical Research (RIKEN), Wako-shi, Saitama, 351-0198, Japan}
\affiliation{Center for Theoretical Physics, Physics Department,
Center for the Study of Complex Systems, The University of
Michigan, Ann Arbor, Michigan 48109-1040, USA}
\date{\today }

\begin{abstract}
We propose a mechanism to interface a transmission line resonator
(TLR) with a nano-mechanical resonator (NAMR) by commonly coupling
them to a charge qubit, a Cooper pair box with a controllable gate
voltage. Integrated in this quantum transducer or simple quantum
network, the charge qubit plays the role of a controllable quantum
node coherently exchanging quantum information between the TLR and
NAMR. With such an interface, a quasi-classical state of the NAMR
can be created by controlling a single-mode classical current in
the TLR. Alternatively, a ``Cooper pair" coherent output through
the transmission line can be driven by a single-mode classical
oscillation of the NAMR.
\end{abstract}


\maketitle

\section{Introduction}
Solid state systems are promising
candidates for novel scalable quantum networks~\cite{Div}.
However, intrinsic features of solid-state-based channels, such as
finite correlation length and environment-induced decoherence, may
limit this scalability. Thus, it is crucial to coherently connect
two or more quantum channels by using suitable quantum nodes.
Coherently interfacing two quantum systems requires a high
fidelity transfer of quantum states between them.

Here we describe a physical mechanism for interfacing a
nanomechanical resonator (NAMR) (see, e.g.,
~\cite{nature1,nature2,sq,schwab,zoller,pra1}) and a
superconducting transmission line resonator (TLR)~\cite{pra1},
i.e., a quantum transducer between mechanical and electrical
signals. With increasing quality factors (e.g., $Q\approx 10^{3}-
10^{5}$) and large eigenfrequencies (e.g., $\omega_b\approx$ MHz -
GHz), NAMRs have been fabricated in the nearly quantum regime and
proposed as candidates for either entangling two JJ qubits~
\cite{armour_PRL_2002,cleland}, or demonstrating progressive
quantum decoherence~\cite{wang}. A superconducting TLR has
recently been demonstrated~\cite{yale_nature_2004} as a quantized
boson mode strongly coupled to a Josephson junction (JJ) charge
qubit~\cite{DML}. Many new possibilities can be explored for
studying the strong interaction between light and macroscopic
quantum systems~(see, e.g., \cite{you01}). In principle, the
quantized boson modes of NAMRs and TLRs can be regarded as quantum
data buses (see, e.g.,~\cite{wei_EPL_2004}). Also, theoretical
proposals have been made for interfacing these with optical
qubits~\cite{tian1,tian2, lukin1}.

Here we investigate the quantum integration of solid-state qubits
and their data buses. In particular, we study how to connect two
very different quantum channels, a mechanical and an electrical,
provided by the NAMR and TLR, through a quantum node implemented
by a Cooper pair box (CPB) or charge qubit. Our system can be
considered the {\it quantum analog of the transducer found in
classical telephones (mechanical vibrations converted into
electrical signals and vice versa)}. Because these three quantum
objects (NAMR, TLR, and CPB) have been respectively realized
experimentally with fundamental frequencies of {\it the same
order}, it is quite natural to expect that they can be effectively
coupled with each other. The physical principle behind our
approach is similar to a theoretical prediction from cavity
QED~\cite{Slesh}: Interacting with a common two-level atom, two
off-resonant boson fields can be effectively entangled and then
the quantum state tomography of a mode can be done with a high
fidelity from the output of another. We similarly use the charge
qubit as an artificial atom to coherently link two kinds of boson
modes, the TLR and the NAMR ones. This quantum-node-induced
interaction is controllable and can be freely switched-on and
-off. A direct TLR and NAMR coupling through the gate voltage is
problematic because the on-chip coupling cannot be easily
controlled.

The physical mechanism, describe below, to prepare the
quasi-classical state of the NAMR has an atomic cavity QED
analogue. Consider an atom located in an optical resonator, and a
classical pump laser also going through the cavity~\cite{Slesh}.
The atom interacts with the cavity field and the laser, and
therefore couples the classical laser to the quantized cavity
field. When the atom is off-resonance with respect to the cavity,
the cavity mode can behave as a forced harmonic oscillator, where
the external force is effectively supplied by the classical laser.
Thus, the coherent state of the cavity mode can be generated and
controlled by the driving laser. This analogy motivates us to
consider an inverse of the above scheme generating the NAMR
coherent state. We set the TLR in a classical oscillation with a
single frequency. This oscillation plays the role of the classical
pump laser in the case of cavity QED. The off-resonant charge
qubit interacts with both the NAMR and TRL, and thus induces an
external force on the NAMR. This force will drive the boson mode
of the NAMR into a coherent state.

\begin{figure}[tbp]
\vspace{-1.5cm} \includegraphics[width=10cm]{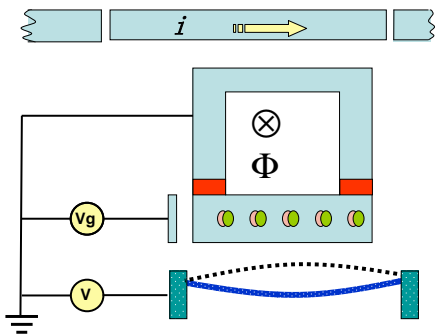}
\vspace{-7.4cm} \caption{Schematic diagram of the coupled system
of a Nanomechanical resonator (NAMR) and Transmission line
resonator (TLR). The TLR is located right above the SQUID
connected to the Cooper-pair box (CPB), so the TLR can produce a
flux threading the loop in the SQUID. The Cooper-pairs are
schematically represented by two small overlapping circles inside
the Cooper-pair box. The NAMR is shown right below the CPB. The
CPB acts as a transducer mediating the interaction between the
NAMR and the TLR.} \label{fig1}
\end{figure}

\section{Model}
The proposed transducer is illustrated
in Fig.~1. A horizontal TLR is fabricated coplanar with a CPB. The
charge state of the CPB can be controlled by the gate voltage
$V_{g}$ applied to the gate capacitor $C_{g}$. The CPB is also
coupled to a large superconductor, the thermal bath, through two
JJs with tunnelling energy $E_{J}$. The SQUID geometry also allows
to apply external magnetic fluxes to control the charge state of
the CPB. A NAMR (at the bottom of Fig.~1) with fundamental
frequency $\omega _{b}$ and mass density $m$ is coupled to the CPB
through the distributed
capacitance $C(x)$, which depends on the displacement
$$
x=\sqrt{\frac{1}{2m\omega _{b}}}(b^{\dag }+b)
$$
quantized by $[b,b^{\dag }]=1$.

Let us assume that the distance fluctuations of the NAMR are much
smaller than the distance $d$ between the NAMR and the CPB. Thus,
the generic formula $C_{d}=\varepsilon A/d$ of the parallel plate
capacitance, with effective area $A$, becomes $C\left( x\right)
\simeq C_{0}(1-x/d)$, where $C_{0}$ is the distributed capacitance
of the NAMR in equilibrium. It is the $x$-dependence of $C\left(
x\right) $ that couples the CPB to the \ NAMR, with free
Hamiltonian $H_{n}=\omega _{b}\,b^{\dag }b$. For small Josephson
junctions, we assume that the equilibrium capacitance $C_{0}$ and
the gate one $C_{g}$ are much less than $C_{J}$. In the
neighborhood of $n_{g}=(C_{g}V_{g}+C_{0}V)/(2e)=1/2,$ the joint
system (CPB and NAMR) can be approximately described by an
effective spin-boson Hamiltonian

\begin{equation}
H_1=\frac{\omega }{2}\sigma _{z}^{\prime }+\omega _{b}\,b^{\dag
}b+\lambda (b^{\dag
}+b)\sigma _{z}^{\prime }-\frac{E_{J}}{2}\cos \left(\frac{\pi \Phi _{x}}{\Phi _{0}%
}\right)\,\sigma _{x}^{\prime }
\end{equation}%
where $\omega =4E_{C}(2n_{g}-1)$, $E_{C}=e^{2}/(2C_{T})$, and
$$
\lambda=\frac{e\,C_{0}\,V}{C_{T}\,d\,\sqrt{2\,m\,\omega _{b}}}.
$$
Above, we have neglected the high frequency term proportional to
$x.$ $C_{T}=C_{J}+C_{g}+C_{0}$ is the total effective capacitance.
The Pauli matrices ($\sigma _{x}^{\prime },\sigma _{y}^{\prime
},\sigma _{z}^{\prime })$ of the quasi-spin are defined with
respect to two isolated charge states, $\left\vert 1\right\rangle
$ and $\left\vert 0\right\rangle $, of the CPB.

The coupling between the CPB and the TLR results from the total
external magnetic flux $\Phi _{x}=\Phi _{C}+\Phi _{q}$ through the
SQUID loop of effective area $S$ $(\approx 1\mu m^{2})$. Here,
$\Phi _{c}$ is a classical flux used to control the Josephson
energy and $\Phi _{q}=S\mu _{0}I_{\max }/(2\pi r)$ is the
quantized flux arising from the quantization of the current $ I $
in the TRL. We assume that the SQUID is placed near the point
where the amplitude of the magnetic field is largest; $r$
$(\approx 10\,\mu m)$ is the distance between the line and the
SQUID, and $\mu _{0}\,(=4\pi \times 10^{-7}{\rm H\,m}^{-1})$ is
the vacuum permeability. The quantized current in the TLR can be
directly obtained from the quantization of the voltage (see, e.g.,
Ref.~\cite{pra1}) through the continuous Kirchhoff's equation
$\partial _{z}I(z,t)=-c\,\partial _{t}V(z,t).$ At the anti-node
$z=L/(2k)$, the quantized current $I(z,t)$ takes its maximum
amplitude $I_{\max }$  to create a quantized flux
\begin{equation}
\Phi _{q}=i\sum_{k}\phi _{k}(a_{k}-a_{k}^{\dag }),\,\, \phi _{k}=\sqrt{\frac{2c%
}{(k\pi )^{3}\,\nu}}\frac{S\mu _{0}L}{2\pi r}.
\end{equation}%
Here, $\nu=1/\sqrt{lc}$\,, with $l$ and $c$ being the inductance
and capacitance per unit length, respectively. At low
temperatures, the qubit
can be only designed to couple a single resonance mode of $\omega _{\mathrm{k%
}}=\omega _{a}$ of the TLR, and then the flux felt by the qubit
becomes $\Phi_{q}=i\phi_a (a-a^{\dag }).$ %

Usually, the quantized flux $\Phi _{q}$ produced by the TRL is not
strong, so that we can expand the Josephson energy to first order
in $\pi \Phi _{q}/\Phi _{0}$. This results in a linear interaction
between the charge qubit and the single mode quantized field.
Namely,
the Josephson coupling $%
V=-E_{J}\cos (\pi \Phi _{x}/\Phi _{0})\sigma _{x}^{\prime }/2$ can
be linearized as
\begin{equation}
 V=i\lambda^{\prime }(a-a^{\dag })\sigma
_{x}^{\prime },\, \lambda^{\prime }=-\frac{E_{J}\pi \phi
_{a}}{2\Phi _{0}}\sin \left( \pi \frac{\Phi _{c}}{2\Phi
_{0}}\right).
\end{equation}
The effective coupling $\lambda^{\prime }$ can be controlled by
the classical external flux $\Phi _{c}$.

Now we choose a new dressed basis (spanned by $\left\vert
e\right\rangle =\cos (\theta/2)\left\vert 0\right\rangle -\sin
(\theta/2)\left\vert 1\right\rangle $ and $\left\vert
g\right\rangle =\sin (\theta /2)\left\vert 0\right\rangle +\cos
(\theta /2)\left\vert 1\right\rangle $) to simplify the above
total Hamiltonian under the rotating-wave approximation. Here, the
mixing angle
$$
\theta =\tan ^{-1}\left[\frac{E_{J}}{\omega}\cos \left( \pi
\frac{\Phi _{x}}{\Phi _{0}}\right)\right]
$$
is calculated with the effective qubit spacing $\epsilon =\sqrt{%
\omega ^{2}+E_{J}^{2}\cos ^{2}\left( \pi \Phi _{x}/\Phi
_{0}\right) }.$ In terms of the the corresponding quasi-spin
(e.g., $\sigma _{x}=|e\rangle \langle g|+|g\rangle \langle e|$),
we obtain the effective Hamiltonian
\begin{eqnarray}
H_2&=&\omega _{a}a^{\dag }a+\omega _{b}b^{\dag }b+\frac{\epsilon
}{2}\sigma
_{z}  \notag  \label{eq24} \\
&&+\lambda_{b}(b\sigma _{+}+\sigma _{-}b^{\dag })+i\lambda_{a}(a\sigma
_{+}-a^{\dag }\sigma _{-}),
\end{eqnarray}%
where two effective coupling constants
$\lambda_{a}=\lambda^{\prime }\cos \theta$ and
$\lambda_{b}=\lambda\sin\theta$ can also be well controlled by the
classical flux.

The coherent interfacing between the TLR and a NAMR implies that
quantum states can be perfectly transferred between them. Let
$\aleph _{T}$ and $\aleph _{N}$ be the Hilbert spaces of the TLR
and NAMR, respectively, and $\left\vert \psi (0)\right\rangle
=\left\vert \psi _{T}(0)\right\rangle \otimes \left\vert \psi
_{N}(0)\right\rangle \in \aleph _{T}$ $\otimes \aleph _{N}$ \ the
initial state of the total system. A generic coherent interfacing
is defined by the factorization of the time evolution $\left\vert
\psi (T)\right\rangle =\left\vert \psi ^{\prime }(T)\right\rangle
\otimes \left\vert \psi _{T}(0)\right\rangle $ at a certain
instance $T$ without any man-made intervention. That is, the local
information carried by $\left\vert \psi _{T}(0)\right\rangle $ in
$\aleph _{T}$ ($\aleph _{N}$) can be perfectly mapped into another
localized in $\aleph _{N}$ ($\aleph _{T}$).

\section{Case I: Quantum information transfer for two degenerate modes}

To explore the essence of the interface between the TLR and the
NAMR, we first consider the degenerate case, i.e., $\omega
_{a}=\omega _{b}$. The dynamics of the degenerate two-mode boson
field coupled to a common two-level atom has been extensively
investigated both analytically and numerically~(see, e.g.,
\cite{2+1a, 2+1b}). It has been proved that, when one mode is in a
coherent state at the initial time $t=0$ and another mode is the
vacuum, an oscillatory net exchange, with a large number of
photons, happens and thus there indeed exists a coherent transfer
of quantum information between them. However, the exchange of
photons between the two modes also displays an amplitude decay and
hence this transfer is not perfect, even without dissipation and
decoherence induced by the environment. In fact, the revivals and
collapses in the boson populations take place over a time-scale
much longer than that of the atomic Rabi oscillations
decay~\cite{2+1a, 2+1b}.

The above ``dynamic collapse" effect can be overcome by
adiabatically eliminating the variables of the CPB in the large
detuning limit:
\begin{equation}
|\Delta |=|\epsilon -\omega _{a}|\gg
G=\sqrt{\lambda_{a}^{2}+\lambda_{b}^{2}}.
\end{equation}
This limit can always be reached, as the effective qubit spacing
$\epsilon $ is adjustable by controlling the gate voltage. Using
the Fr\"{o}hlich-Nakajima transformation~\cite{Fro,Nakajima},
\begin{eqnarray}
H_S&=&\exp(-S)H_2\exp(S)\nonumber\\
&=&H_2+[H_2,S]+\frac{1}{2}[[H_2,S],S]+...,
\end{eqnarray}
with
$$S=G(A\sigma_+-A^\dagger\sigma_-)/\Delta,\,\,A=b\cos \beta +ia\sin \beta,$$  we obtain an effective Hamiltonian
\begin{equation}
H_3\simeq \omega _{a}(A^{\dag }A+B^{\dag }B)+(\frac{\epsilon}{2}
-\delta )\sigma _{z}-\delta A^{\dag }A\sigma _{z},
\end{equation}%
approximated to first-order in the small quantity $G/\Delta$.
Here, $\delta =G^{2}/\Delta $ is the Stark shift and $\beta
=\arctan (\lambda_{a}/\lambda_{b})$. Besides $A$, we have
introduced another normal-mode
$$
B=b\sin \beta -ia\cos \beta.
$$
The above effective Hamiltonian shows that, when the charge qubit
can adiabatically remain in the ground state $\left\vert
0\right\rangle $, the two boson modes $a$ and $b$ evolve according
to two normal modes $A$ and $B$ with a frequency difference
$\delta $. The non-zero frequency difference $\delta $ between the
modes $A$ and $B$ results in the coherent exchange of these boson
numbers. In fact, on account of the exact solution $A(t)$ and
$B(t)$ of eigenmodes, the Heisenberg
equation for the natural modes can be solved as%
\begin{eqnarray}
a(t) &=&a(0)F_{1}(t)+b(0)K(t),  \notag \\
b(t) &=&b(0)F_{2}(t)-a(0)K(t),
\end{eqnarray}%
where the time-dependent coefficients are
\begin{eqnarray}
F_{k}(t) &=& \left [\cos \left(\frac{\delta
t}{2}\right)+i(-1)^{k}\cos (2\beta
)\sin \left(\frac{\delta t}{2}\right) \right]\exp(-i\Theta t),  \notag \\
K(t) &=&\sin (2\beta )\sin \left(\frac{\delta
t}{2}\right)\exp(-i\Theta t),
\end{eqnarray}%
for $\Theta =\omega _{a}-\delta /2$, and $k=1,2.$

Having the explicit expressions for the Heisenberg operators $a$
and $b$, the algebraic technique developed in \cite{gao} can be
used to explicitly construct the wave function of the NAMR-TRL
interfacing system. When the initial state of the joint system
(NAMR and TRL) is $|\Psi (0)\rangle =|n\rangle \otimes |0\rangle
,$ the wave function at time $t$ becomes $|\Psi (t)\rangle
=[a^{\dagger }(-t)]^{n}|0\rangle/\sqrt{n!}$, or
\begin{equation}
|\Psi (t)\rangle =\frac{1}{\sqrt{n!}}[a^{\dagger }(0)F_{k}^{\ast
}(-t)+b^{\dagger }(0)K^{\ast }(-t)]^{n}|0\rangle .
\end{equation}%
To realize a perfect interface between the NAMR and the TRL, we
need to consider whether $a(0)$ can oscillate into $b(0)$ in a
certain instance, and vice-versa. In Fig.2,
\begin{figure}[tbp]
\vspace{-3.5cm} \includegraphics[width=8.8cm]{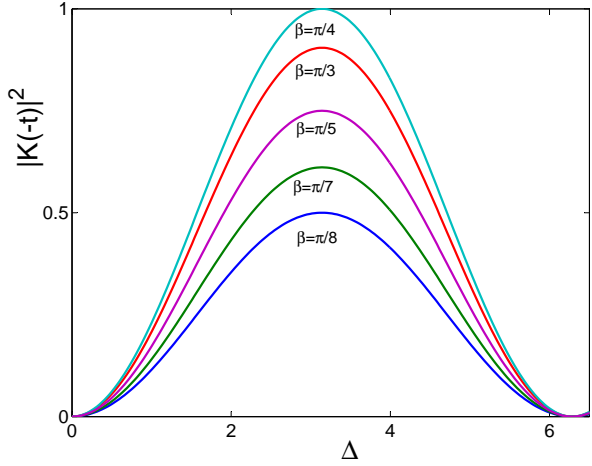}
\vspace{-2.8cm} \caption{Parameter $|K(-t)|^{2}$ changes with
time-dependent variable $\Delta=t\,\delta$ for different $\beta$
values.} \label{fig1}
\end{figure}
we draw the curves of $|K(-t)|$ changing with time $t$ for
different parameters $\beta $. For $\beta =\pi /4$, one can easily
see that $|K(-t)|$ can reach unity while $|F_{k}(-t)|$ vanishes.
This implies that a perfect exchange of quantum states can be
implemented between the NAMR and the TRL. Mathematically, when
$\beta =\pi /4,\,F_{k}(t)$ and $K(t)$ define two complementary
oscillations with amplitudes ranging from $0$ to $1$. The simple
amplitude complementary relation
\begin{equation}
|F_{k}(t)|^{2}+|K(t)|^{2}=1
\end{equation}
and the same phase factor means a perfect transfer of quantum
states. Physically, $\beta =\pi /4$ means that the effective
couplings $\lambda _{a}$ and $\lambda _{b}$, of the NAMR and the
TLR, are the same. Indeed, we can realize the perfect transfer of
quantum information at the moments $t=(2m+1)/\delta $ for
$a(-t)=b(0)\exp (i\omega _{a}t)$, i.e., the wave function can be
factorized into
$$\sum
c_{n}|n\rangle \otimes |0\rangle \rightarrow W\sum c_{n}|0\rangle
\otimes |n\rangle $$
by a known unitary transformation
$$W={\rm diag}\,\{\exp (i\omega _{a}t),\exp (i2\omega _{a}t),...,\exp
(in\omega _{a}t)\},$$ which is independent of the initial state.

\section{Case II: Quantum information transfer for two non-degenerate modes}

In the degenerate case we have demonstrated the perfect
transmission of quantum states between the NAMR and TLR by
connecting them via a charge qubit. In principle, it is also
possible to perform quantum information transfer between two
non-degenerate modes. In fact, the model of two non-degenerate
modes coupled to a two-level system can be solved exactly, and the
phenomenon of rapid-collapse and revival could be
shown~\cite{jmo01}. However, it is convenient to adiabatically
eliminate the connecting qubit for directly transferring quantum
states between the two non-degenerate modes.

Again, we assume that the large detuning condition is still
satisfied. To directly connect the two non-degenerate modes by
adiabatically eliminating the qubit, we introduce an
anti-Hermitian operator
\begin{equation}
W=-i\frac{\lambda_a}{\Lambda}(a\sigma_++a^\dagger\sigma_-)-\frac{\lambda_b}{\Lambda}(b\sigma_+-b^\dagger\sigma_-),
\end{equation}
to perform the Fr\"{o}hlich-Nakajima transformation on $H_2$ and
obtain the following effective Hamiltonian
\begin{eqnarray}
H_4&\simeq& \omega _{a}a^{\dag }a+\omega _{b}b^{\dag
}b+\frac{\Omega }{2}\sigma _{z}
+\left(\frac{\lambda_a^2}{\Lambda}a^\dagger
a+\frac{\lambda_b^2}{\Lambda}b^\dagger b\right)\sigma_z\notag
\\&&+i\frac{\lambda_a\lambda_b}{\Lambda}(ab^\dagger-a^\dagger
b)\sigma_z.
\end{eqnarray}%
with $\Omega=\epsilon+(\lambda_a^2+\lambda_b^2)/\Lambda$. The
detuning
$$\Lambda =-\omega _{a}-\omega _{b}+\Omega$$ is set to satisfy the
conditions:
\begin{equation}
\lambda_a,\,\lambda_b \gg \Lambda.
\end{equation}
The anti-Hermitian operator $W$ satisfies the condition
\begin{equation}
H_2-H_0+[H_0,W]=0,\,\,\,H_0=\omega_aa^\dagger a+\omega_b b^\dagger
b+\frac{\epsilon}{2}\sigma_z,
\end{equation}
which means that the first-order correction vanishes and the above
approximation is second-order perturbation.

Without loss of generality, the charge qubit could be
adiabatically fixed in the ground state $|0\rangle$. As a
consequence, the dynamics of this two-boson system can be
described by
\begin{eqnarray}
H'_4\,=\,\Gamma_0\,\hat{N}+\Gamma_2\,\hat{J}_y+\Gamma_3\,\hat{J}_z
\end{eqnarray}
with
\begin{eqnarray*}
\Gamma_0&=&\frac{\omega_a+\omega_b}{2}+\frac{\lambda_a^2+\lambda_b^2}{2\Lambda},\nonumber\\
\Gamma_2&=&\frac{2\lambda_a\lambda_b}{\Lambda},\nonumber\\
\Gamma_3&=&\omega_a-\omega_b+\frac{\lambda_a^2-\lambda_b^2}{\Lambda},\nonumber\\
\end{eqnarray*}
and $\hat{N}=b^{+}b+a^{+}a$. Angular momentum operators $\hat{J}_l
\,(l=x,y,z)$, defined by the following Jordan-Schwinger
realizations
\begin{eqnarray}
\hat{J}_{x}&=&\frac{b^{+}a+a^{+}b}{2},\nonumber\\
\hat{J}_{y}&=&\frac{i(b^{+}a-a^{+}b)}{2},\\
\hat{J}_{z}&=&\frac{a^{+}a-b^{+}b}{2},\nonumber
\end{eqnarray}
form a
dynamic $SO(3)$ algebra:
\begin{equation}
[\hat{J}_z,\hat{J}_x]=i\hat{J}_y,\,\,[\hat{J}_y,\hat{J}_z]=i\hat{J}_x,\,\,[\hat{J}_x,\hat{J}_y]=i\hat{J}_z.
\end{equation}
Obviously, $\hat{N}$ commutes with the operators $\hat{J}_z$ and
$\hat{J}_y$. This implies that the Hamiltonian $H'_4$ describes a
high-spin precession in an external ``magnetic field"
$B=(0,\Gamma_2,\Gamma_3)$, and thus is exactly
solvable~\cite{sun-z}. The corresponding time-evolution operator
is
\begin{eqnarray}
\hat{U}(t)&=&\exp(-\frac{i}{\hbar}\Gamma_0\,\hat{N}t)\,\exp(-\frac{i}{\hbar}\tilde{H}t),\nonumber\\
\tilde{H}&=&\tilde{\Gamma}\exp(i\beta\,\hat{J}_x)\,\hat{J}_z\,\exp(-i\beta\,\hat{J}_x),
\end{eqnarray}
with $\tilde{\Gamma}=\sqrt{\Gamma_2^2+\Gamma_3^2}$, and
$\tan\beta=\Gamma_2/\Gamma_3$.

The above dynamics can be used to achieve the transfer of an
arbitrary quantum state between the two non-degenerate modes. As a
simple example, we discuss how to transfer a single-phonon state
$|1_b\rangle$ from the NAMR to the TLR, whose initial state is the
vacuum state $|0_a\rangle$. The initial state of this two-mode
system is $|\psi (0)\rangle =|0_a,1_b\rangle$. The wave function
at time $t$ reads
\begin{eqnarray}
|\psi(t)\rangle&=&\hat{U}(t)\,|\psi(0)\rangle\nonumber\\
&=&\left[\cos\left(\frac{\tilde{\Gamma\,
t}}{2\hbar}\right)+i\cos\beta\sin\left(\frac{\tilde{\Gamma\,
t}}{2\hbar}\right)\right]|0_a,1_b\rangle\nonumber\\
&-&\sin\beta\sin\left(\frac{\tilde{\Gamma\,
t}}{2\hbar}\right)|1_a,0_b\rangle.
\end{eqnarray}
If $\Gamma_3=0$, a perfect transfer of quantum information is
obtained by setting the duration as $\sin[\tilde{\Gamma
t}/(2\hbar)]=\pm 1$. For the generic case, $\Gamma_3\neq 0$, a
projective measurement $\hat{P}_b=|0_b\rangle\langle 0_b|$ acting
on the NAMR is required for projecting the TLR collapse to the
desirable state $|1_a\rangle$. The rate of this transfer is
\begin{equation}
P(t)=\sin^2\beta\sin^2\left(\frac{\tilde{\Gamma\,
t}}{2\hbar}\right),
\end{equation}
with the maximal value $\sin^2\beta$ corresponding to the duration
$\sin[\tilde{\Gamma t}/(2\hbar)]=\pm 1$.

\section{Quasi--classical state of the Nano-mechanical resonator}

Above, we have discussed how to transfer a quantum state from the
NAMR to the TLR. Now, we investigate the preparation of a
quasi-classical state of the NAMR, driven by a classical current
input from the TLR. Adiabatically eliminating the connecting qubit
results in an indirectly coupling between the TLR and the NAMR.
Via such a virtual process, the current in TLR produces an
effective linear force acting on the NAMR mode. This force causes
a quasi-classical deformation of the NAMR. Therefore, a coherent
state, which is described by a displaced Gaussian wave packet in
the spatial position, can be generated in the NAMR mode.

For this goal, we treat the driving current classically by the
Bogliubov approximation, that replaces the above annihilation and
creation
operators $a$ and $a^{\dag }$ by the complex amplitudes $\xi=\mu \exp [-i\varphi ]$ and $%
\xi^*=\mu \exp [i\varphi ]$ respectively, where the real numbers $\mu $ and $%
\varphi $ are the amplitude and phase of the classical current,
respectively. We assume, like in the previous section, that the
large detuning condition is still satisfied. Thus, one can
adiabatically eliminate the connected qubit and obtain a
semi-classical Hamiltonian
\begin{equation}
H_{e}=\Omega_{b}b^{\dag }b+i\frac{\Gamma_2}{2}\mu (e^{-i\varphi
}b^{\dag }-be^{i\varphi }),
\end{equation}%
with $\Omega_{b}=\omega_b+\lambda_b^2/\Lambda$. This Hamiltonian
drives the NAMR to evolve from a vacuum state $|0\rangle $ to the
coherent state
\begin{equation}
|z(t)\rangle
=\exp[-|z(t)|^2]\,\sum_{n=0}\frac{[z(t)]^n}{\sqrt{n!}}|n\rangle,
\end{equation}
with
$$z(t)=-i\frac{\Gamma_2\,\xi}{2\,\Omega_b}[1-\exp(-i\Omega_b\,
t/\hbar)].$$
The above coherent state (23) corresponds to a coherent
oscillation in a normal mode of the NAMR. The square of the
coherent state amplitude represents the population rate of the
boson excitation in the transmission line.

To this end, we require a classical TLR current in a single mode,
which plays a similar role as the classical pump laser in optical
masers. While switching on the coupling with the off-resonance
charge qubit for a while, the charge qubit results in a virtual
process as an effective linear force on a NAMR mode. It thus
causes a quasi-classical deformation of the NAMR, described by a
coherent state, which is a displaced Gaussian wave packet in the
spatial position. This physical mechanism is very similar to that
of the pulsed atomic laser~\cite{BEC}.

Even without adiabatic elimination for large detuning, we can
still achieve the same qualitative conclusion for the state
preparation. In the two cases: (a) $\omega _{b}=\epsilon $, and
(b) $\omega _{a}=\omega _{b}$, the achieved semi-classical
Hamiltonian
\begin{equation}
H_{c}=\frac{\epsilon}{2}\sigma _{z}+\omega_b b^\dagger
b+[(\lambda_{b}b +i\lambda_{a}\xi)\sigma _{+}+h.c]
\end{equation}
describes a driven Jaynes-Commings model. Now, we can uniquely
deal with
both cases as follows. If we define the displaced boson operator
$$
B^\prime=b+i\lambda_{a}\xi,
$$
$H_{c}$ becomes the standard Jaynes-Commings Hamiltonian with
interaction $\lambda_{b}(B^\prime\sigma _{+}+h.c)$, but its ground
state experiences a symmetry-breaking. Let $|n(z)\rangle
=D(-z)|n\rangle $ be the displaced Fock state defined by the
coherent state generator $D(z)=\exp (zb^{\dag }-z^{\ast }b).$ The
ground state of the NAMR-CPB composite system is just a product
state $|\alpha =i\lambda_{a}\xi \rangle \otimes |g\rangle $,
basically consisting of a coherent state of the NAMR. This simple
observation reveals that the charge-qubit-based preparation of the
quasi-classical state of the NAMR is robust.

\section{Concluding remarks}

In summary, we propose a mechanism to interface a transmission
line resonator (TLR) with a nano-mechanical resonator (NAMR) by
commonly coupling them to a charge qubit, a Cooper pair box with a
controllable gate voltage. Integrated in this quantum transducer
or simple quantum network, the charge qubit plays the role of a
controllable quantum node coherently exchanging quantum
information between the boson modes of the TLR and NAMR. We have
shown that quantum information can be transferred between these
two, both degenerate and non-degenerate, boson modes.
Also, with such an interface, a quasi-classical state of the NAMR
can be created by controlling a single-mode classical current in
the TLR. Alternatively, a ``Cooper pair" coherent output through
the transmission line can be driven by a single-mode classical
oscillation of the NAMR.

This work was supported in part by the National Security Agency
(NSA) and Advanced Research and Development Activity (ARDA) under
Air Force Office of Research (AFOSR) contract number
F49620-02-1-0334, and by the National Science Foundation grant No.
EIA-0130383. The work of C.P.S is also partially supported by the
NSFC, and FRPC with No. 2001CB309310.

\end{document}